\documentclass[prl,showpacs,twocolumn,amssymb]{revtex4}
\begin{document}

\title{Anisotropic Node Removal in d-wave Superconductors under Magnetic Field}
\author{Efrain J. Ferrer}
\author{Vivian de la Incera}
\affiliation{Physics Department, State University of New York at
Fredonia, Fredonia, NY 14063, USA}

\begin{abstract}
A phenomenological model that considers different secondary
$id_{xy}$ gap amplitudes and quasiparticle effective charges for
each nodal direction, is proposed to explain the observed
anisotropic node removal by a magnetic field in high-$T_{c}$
cuprates. Two independent parity-breaking condensates
$\langle\overline{\Psi_{i}}\bigwedge{\Psi}_{i}\rangle$ develop,
implying the induction of a magnetic moment per each nodal
direction. The model outcomes are in agreement with the
experimentally found relation $\Delta_{i}\sim \sqrt{B}$. The
secondary gap vanishes through a second-order phase transition at
a critical temperature whose value for underdoped nodal-oriented
YBCO is estimated to be $\sim 6K$ for a field of $15T$.
\pacs{74.20.Rp; 74.20.-z; 74.72.-h}
\end{abstract}
\maketitle

Thanks to a wealth of experiments and theoretical studies on high
$T_{c}$ cuprates, it has become clear that any viable theory of
high-$T_{c}$ superconductivity should incorporate, beside the
superconducting and normal phases, several other ordered phases
that may be externally induced or spontaneously generated, and can
either coexist or compete.

The idea of the existence of competing ground states and quantum
critical points in high-$T_{c}$ cuprates was used by Votja, Zhang
and Sachdev \cite{votja1,votja2} to explain an observed
anomalously large damping rate of quasiparticles near the gap
nodes of the $Bi_{2} Sr_{2}CaCu_{2}O_{8+\delta}$ superconductor.
The quasiparticle damping rate, studied in photoemission
spectroscopy (ARPES) \cite{valla} and optical conductivity
\cite{corson} experiments, revealed a linear in temperature
dependence, in contrast to the cubic dependence expected in the
conventional d-wave picture \cite{quinlan}. This anomalous
enhancement of the rate was considered in Ref.
\cite{votja1,votja2} to be associated to a quantum critical
behavior and interpreted as the result of the fluctuations of a
new, secondary order parameter that removed the quasiparticle
nodes. Then, a phenomenological theory of nodal quasiparticles
coupled to the fluctuations of a secondary ($is$ or $id_{xy}$)
pairing order parameter was proposed, and evidence of a quantum
critical point in the superconducting phase was provided within a
perturbative $\epsilon$-expansion in Ref. \cite{votja2}. This
quantum critical point was later established beyond perturbation
theory by Khveshchenko and Passke \cite{khvesh}, finding a damping
rate $\sim T$ at high temperatures.

The development of a new order parameter has been also found in
tunneling experiments with $YBa_{2}Cu_{3}O_{7-x}$ (YBCO)
\cite{b-tunneling}. In these studies, a splitting of the zero bias
conductance peak (ZBCP) is observed, when a magnetic field is
perpendicularly applied to the $CuO_{2}$ planes of the YBCO films.
A ZBCP is associated to the presence of nodes, as it reflects the
existence of Andreev surface bound states at and near the Fermi
surface \cite{hu}. Hence, the splitting of the ZBCP may be an
indication of the appearance of a secondary gap, and therefore of
node removal by the magnetic field. In addition to the field
induced splitting of the ZBCP, a spontaneous -probably associated
to the doping level- splitting was also seen in
\cite{b-tunneling}. These findings motivated Dagan and Deustscher
\cite{dagan1} to perform a comprehensive study of the doping
dependence of the ZBCP splitting in YBCO samples with and without
external magnetic field. These authors concluded that above
certain doping level near optimum doping, a spontaneous ZBCP
splitting occurs, which increases with doping. It also increases
with magnetic field applied along the c-axis. In underdoped
samples the magnetic field could induce the ZBCP splitting, but no
spontaneous splitting was observed.

Although the ZBCP splitting by the magnetic field may be
considered as due to node removal, a different explanation, based
on Doppler shift of the Andreev bound states has been also offered
\cite{fogels}. However, a new set of experiments \cite{dagan2} has
definitely favored node removal in the case of underdoped
materials, measured in decreasing fields to eliminate hysteresis
effects. In these experiments a large ZBCP splitting was clearly
observed in films with thickness smaller than the London
penetration depth. That is, under conditions where the Doppler
shift effect is negligible. Another important finding of Ref.
\cite{dagan2} is that the splitting is very anisotropic. Only for
samples having (1,1,0) orientation, the splitting was substantial.
The authors of \cite{dagan2} noticed that the experimental data
obtained for the (1,1,0) oriented samples were in good agreement
with the predictions made by Laughlin \cite{laugh} in a model he
had proposed some years ago to explain the behavior of the thermal
conductivity of cuprates in a magnetic field \cite{krisha}.
However, they also called attention that at present there is no
theoretical model that predicts the observed anisotropy. Laughlin
model, which is based on a transition from a pure d-wave order
parameter to a parity and time-reversal symmetry violating
$d_{x^{2}-y^{2}}+id_{xy}$ ground state, removes the nodes of both
nodal directions.

It should be mentioned that the behavior of the thermal
conductivity of cuprates with a magnetic field has proven itself
to be anisotropic too. The hallmark of Krishana \textit{et al.}
effect \cite{krisha}, namely, the decreasing of the thermal
conductivity with the magnetic field, followed by a kink and a
plateau region, is indeed observed in only one of the two nodal
orientations \cite{ando1}. This striking result was revealed,
after a long debate, by careful experiments realized by Ando
\textit{et al.} \cite{ando1}. Anisotropic in-plane transport
properties of cuprates have been also observed by Ando and
coworkers in measurements of the electrical resistivity of
underdoped samples \cite{ando2} and of the magnetoresistance of
lightly doped ones \cite{ando3}. These experimental findings have
indicated that the anisotropy between the nodal directions is
related to the presence of charge stripes along the nodal
directions of these materials. Thus, the electrons organize in an
anisotropic macroscopic state, characterized by stripes that
depend on the temperature and doping.

In this paper we propose a phenomenological model for the
low-energy quasiparticle excitations at the nodes of a
superconductor on which both magnetic field and doping can induce
the transition from the regular d-wave pairing to
$d_{x^{2}-y^{2}}+id_{xy}$- or $d_{x^{2}-y^{2}}+is$-pairings. The
novel feature of the present model is that it incorporates the
observed anisotropy of the nodal directions, so it can describe
the anisotropic field-induced ZBCP splitting phenomenon
\cite{dagan1,ando2}. Two main new ideas have been implemented
here: (1) to consider the induction of a secondary order parameter
with $id_{xy}$ (or $is$) symmetry that may have different
amplitudes along the two nodal directions, and, (2) inspired by
the above mentioned anisotropic charge distribution along the
nodal directions, to allow for different effective charges of the
quasiparticle excitations about these directions.

Our treatment will follow the standard procedure based on the
rotationally covariant Nambu representation, employed in Refs.
\cite{votja1,votja2} and \cite{khvesh}, to describe the
quasiparticles in terms of two species of Nambu bispinors. Then,
the bispinors are combined in four-component Dirac spinors, which
permit to write the action in a familiar field theory form.
Following Lee and Wen \cite{lee}, we assume that the spectrum of
the quasiparticle excitations in the superconductor is given by
$E(k)=\sqrt{\epsilon(\textbf{k})+\Delta^{2}(\textbf{k})}$. Here
$\epsilon(\textbf{k})=2t(\cos(bk_{x})+\cos(bk_{y}))$, as usual.
For the proposed complex order parameter $\langle
c_{\textbf{k}\uparrow}c_{\textbf{-k}\downarrow}\rangle$ =
$\Delta(\textbf{k})$, however, we propose a secondary $id_{xy}$
component characterized by two amplitudes $\Delta_{ixy}$, ($i=1,2$
), one for each nodal direction, given by
$\Delta(k)=\Delta_{0}(\cos(bk_{x})-\cos(bk_{y})
)+i\Delta_{1xy}\sin(bk_{x})sin(bk_{y})
[sin^{2}(\frac{b}{2}(k_{x}-k_{1x}))sin^{2}(\frac{b}{2}(k_{y}-k_{1y})+
sin^{2}(\frac{b}{2}(k_{x}+k_{1x}))sin^{2}(\frac{b}{2}(k_{y}+k_{1y}))]+
(1\leftrightarrow2)$, where $\pm k_{i}=\pm( k_{F},\pm k_{F})$ are
the momentum of the nodes along the two nodal directions. A
similar order parameter can be used to produce $is$ secondary
pairing, by substituting the factor $\sin(bk_{x})sin(bk_{y})$ by
$1$ in $\Delta(k)$. From now on we concentrate our discussion on
the $id_{xy}$, as the results can be easily extended to the $is$
case. Our guidance to propose such a form of the secondary order
parameter has been to use a basis function that transforms under
the symmetry group of the superconductor $C_{4v}\times Z_{2}$
\cite{annett} in the same way as the usual $id_{xy}$  basis
function.

Rotating the coordinate system by $\pi/4$ and expanding about each
nodal point up to linear order, one can write the low energy
effective action of the quasiparticle excitations around the nodes
as the sum of a fermion kinetic term
\begin{eqnarray}
S_{\varphi }=i\sum\limits_{\alpha =\uparrow }^{\downarrow }\int d^{3}x[%
\overline{\varphi }_{1\alpha }\left( \gamma _{0}\partial
^{0}+v_{F}\gamma _{x}\partial ^{x}+v_{\Delta }\gamma _{y}\partial
^{y}\right) \varphi _{1\alpha }\nonumber\\*+(\varphi _{1\alpha
}\rightarrow \widetilde{\varphi }_{2\alpha },x\longleftrightarrow
y)],
\end{eqnarray}
a Higgs-Yukawa (HY) term
\begin{eqnarray}
S_{\varphi \phi }=\int d^{3}x[g\phi _{1}\sum\limits_{\alpha
=\uparrow }^{\downarrow }\overline{\varphi }_{1\alpha }\varphi
_{1\alpha }+g\phi _{2}\sum\limits_{\alpha =\uparrow }^{\downarrow
}\overline{\widetilde{\varphi }}_{2\alpha }\widetilde{\varphi
}_{2\alpha }],
\end{eqnarray}
and a scalar field action, whose generic form can be found, after
integrating out high-energy fermion modes, to be
\begin{eqnarray}
S_{\phi }=\int d^{3}x[\frac{1}{2\overline{c}^{2}}\left( \partial
_{0}\phi _{1}\right) ^{2}+\frac{1}{2}\left( \nabla \phi
_{1}\right) ^{2}-m^{2}\phi _{1}^{2}-\lambda \frac{\phi
_{1}^{4}}{4!}\nonumber\\
+\frac{1}{2\overline{c}^{2}}\left( \partial _{0}\phi _{2}\right)
^{2}+\frac{1}{2}\left( \nabla \phi _{2}\right) ^{2}-m^{2}\phi
_{2}^{2}-\lambda \frac{\phi _{2}^{4}}{4!}]
\end{eqnarray}
In the above equations $m,\lambda,g$ and $\overline{c}$ are
undetermined parameters to be fixed by the experiment,
$v_{F}=2\sqrt{2}tb$  and $v_{\Delta }=\frac{b\Delta
_{0}}{\sqrt{2}}$ are the characteristics velocities of the d-wave
superconductor, and $g\phi _{1}\equiv \Delta _{1xy}$ and $g\phi
_{2}\equiv \Delta _{2xy}.$ We are using the irreducible
representation $\gamma _{\mu }=(\tau _{2},i\tau _{1},-i\tau
_{3})$, $\mu =0,1,2$, of the (2+1)-dimensional Dirac algebra,
where $\tau _{i}$ are the Pauli matrices. The quasiparticle states
near the pairs of the opposite nodes have been described in terms
of the bispinors $\varphi _{i\alpha }=\left( c_{\alpha }\left(
k_{i}\right) ,\varepsilon _{\alpha \beta }c_{\alpha }^{\dagger
}\left( -k_{i}\right) \right) $, with $\widetilde{\varphi
}_{2\alpha }=\frac{\tau _{1}+\tau _{3}}{\sqrt{2}}\varphi _{2\alpha
}$, and we introduced the conjugates $\overline{\varphi }_{1\alpha
}=\varphi _{1\alpha }^{\dagger }\gamma _{0}$,
$\overline{\widetilde{\varphi }}_{_{2\alpha }}=$
$\widetilde{\varphi }_{2\alpha }^{\dagger }\gamma _{0}.$ The index
$\alpha =\uparrow ,\downarrow $ denotes the two spin components,
and $\varepsilon _{\alpha \beta }$ is an antisymmetric tensor with
$\varepsilon _{\uparrow \downarrow }=1.$

To make the fermion sector of the action explicitly Lorentz
invariant, we can make the variable change $x\rightarrow
\sqrt{\frac{v_{F}}{v_{\Delta }}}x,$ $y\rightarrow
\sqrt{\frac{v_{\Delta }}{v_{F}}}y$, in the terms with the $\varphi
_{1\alpha }$ field, and $x\rightarrow \sqrt{\frac{v_{\Delta
}}{v_{F}}}x,$ $y\rightarrow \sqrt{\frac{v_{F}}{v_{\Delta }}}y$, in
those with the field $\widetilde{\varphi }_{2\alpha }$. In this
way, $v=\sqrt{v_{F}v_{\Delta }}$ plays the role of the speed of
light. From now on, for simplicity, we set all the velocities to
1.

At this point it is convenient to consider a reducible 4$\times $4
representation of the Dirac algebra by introducing the matrices
$\Gamma _{\mu }=diag(\gamma _{\mu },-\gamma _{\mu }),$ and
combining the bispinors of each nodal direction in four-component
Dirac spinors, to form two four-spinors of different "flavors"
$i=1,2$ defined as $\Psi _{1}=(\varphi _{1\uparrow },\varphi
_{1\downarrow }),$ $\Psi _{2}=(\widetilde{\varphi }_{2\uparrow
},\widetilde{\varphi }_{2\downarrow })$ with conjugate
$\overline{\Psi }_{i}=\Psi _{i}^{\dagger }\Gamma _{0}.$ Notice
that the $i^{th}$ Dirac spinor involves the fields of the $i^{th}$
nodal direction only. In the $4\times 4$ representation the action
becomes
\begin{eqnarray}\label{fouraction}
S=\int d^{3}x\sum_{i=1}^{2}[\overline{\Psi}_{i}\Gamma_{\mu}
\partial^{\mu}\Psi_{i}+g\phi_{i}\overline{\Psi}_{i}\Lambda
\Psi_{i}\nonumber\\
+\frac{1}{2}\left(\partial_{\mu }\phi
_{i}\right)^{2}-m^{2}\phi_{i}^{2}-\lambda \frac{\phi_{i}^{4}}{4!}]
\end{eqnarray}
where
$\Lambda=i\Gamma_{0}\Gamma_{1}\Gamma_{2}=diag(I_{2},-I_{2}).$ The
action (\ref{fouraction}) is invariant under parity
transformations
\begin{equation}\label{symtranf}
 \left(x\rightarrow -x,y\rightarrow y\right),\phi _{i}\rightarrow
-\phi _{i},\Psi _{i}\rightarrow \Gamma _{P}\Psi _{i}
\end{equation}
with $\Gamma _{P}=\tau_{1}\otimes\tau_{1}$.

Let us now switch on a magnetic field along the c-axis. As
mentioned above, an important point of the proposed model is to
consider that the effective charges of the two nodal directions
are different. Thus, the magnetic field couples with different
strengths $e_{1}$ and $e_{2}$ to fields $\Psi _{1}$ and  $\Psi
_{2}$ respectively. Then, in the presence of the magnetic field
the action (\ref{fouraction}) takes the form
\begin{eqnarray}\label{baction}
S_{B}=\int d^{3}x\sum_{i=1}^{2}[\overline{\Psi }_{i}\Gamma _{\mu
}D_{i}^{\mu }\Psi _{i}+g\phi _{i}\overline{\Psi }_{i}\Lambda \Psi
_{i}\nonumber\\+\frac{1}{2}\left(
\partial _{\mu }\phi _{i}\right) ^{2}-m^{2}\phi _{i}^{2}-\lambda
\frac{\phi _{i}^{4}}{4!}]
\end{eqnarray}
where $D_{i}^{\mu }=\partial _{\mu }-ie_{i}A_{\mu }^{ext},$ and we
take a gauge on which $A_{\mu }^{ext}=\left( 0,0,Bx_{1}\right).$
If due to the magnetic field the scalar field acquires nonzero
vacuum expectation value (vev) $\widehat{\phi }_{i}$ (induced
secondary energy gap), the corresponding $i^{th}$-parity symmetry
is broken. To explore such a possibility it is needed to minimize
the free energy of the system with respect to $\widehat{\phi
}_{i}$. In the lowest approximation, it gets contributions from
all tadpole diagrams of the effective theory. At finite
temperature the minimum equation is
\begin{eqnarray}\label{gapeq1}
2m^{2}\phi+\frac{\lambda }{6}\phi _{i}^{3}=\nonumber\\-\frac{\hbar
}{8\pi ^{2}\beta }\lambda \phi _{i}\sum_{n=-\infty }^{\infty }\int
d^{2}k\frac{1}{\left( \frac{2n\pi }{\beta }\right)
^{2}+k^{2}+m^{2}+\lambda \frac{\phi _{i}^{2}}{2}}
\nonumber\\+\hbar \frac{2g^{2}\left| e_{i}B\right| \phi _{i}}{\pi
\beta }\sum_{l=0}^{\infty }\sum_{n=-\infty }^{\infty
}\frac{1}{\left( \frac{(2n+1)\pi }{\beta }\right) ^{2}+2\left|
e_{i}B\right| l+g^{2}\phi _{i}^{2}}
\end{eqnarray}
where $\beta^{-1}=T$, and, to simplify the notation, we dropped
the use of the "hat" in the scalar vev's.

Since this is an effective low-energy theory of quasiparticle
excitations around the nodes, the low-energy domain will be
limited by the induced gap. Therefore, the divergent integrals
should have cutoff $\Delta_{i}$. Moreover, it is natural to expect
a magnetically induced gap smaller than the field generating it.
As a consequence, the fermion infrared dynamics will be dominated
by the lowest Landau level, and the leading contribution to the
fermion tadpole, second term in the \textit{RHS} of
(\ref{gapeq1}), will come from $l=0$. Thereby, after doing the
temperature sums, we obtain
\begin{widetext}
\begin{equation}\label{gapeq2}
\frac{1}{3g\alpha^{2}}\Delta_{i}
^{3}+\frac{\hbar}{8\pi\alpha}g\Delta_{i} ^{2}+\frac{\hbar
}{2\pi}\frac{g\Delta_{i} }{\alpha^{2} }\sqrt{\frac{\Delta_{i}
^{2}}{\alpha ^{2}}+T^{2}}e^{(-\frac{\Delta_{i} }{2\alpha
T})}-\hbar \frac{g\left| e_{i}B\right| }{\pi }\tanh \left(
\frac{\Delta_{i} }{2T }\right)=0
\end{equation}
\end{widetext}
In (\ref{gapeq2}) we took $m^{2}=0$, and introduced the notation
$\alpha\equiv\sqrt{\frac{2g^{2}}{\lambda}}$. Eq. (\ref{gapeq2})
can have different field-dependent solutions, for different ranges
of $\lambda$ and $g$. However, to describe the experimental
findings, we need to consider the parameter range that leads to
$\Delta_{i}\sim \sqrt{|e_{i}B|}$ \cite{dagan2}. Such a solution
can be found neglecting the first term in the \textit{LHS} of
(\ref{gapeq2}), which corresponds to neglect the contribution of
the tree-level $\lambda\phi^{4}$ term in the free-energy. In this
way, the zero-temperature minimum results
\begin{equation}\label{zero-T-delta}
  \Delta^{(0)}_{i}=2\sqrt{\alpha}\sqrt{2|e_{i}B|.}
\end{equation}
Experiments with YBCO samples gives the ratio
$\frac{\Delta^{(0)}_{i}}{\sqrt{B}}= 1.1 meV/T^{1/2}$ in data taken
in decreasing fields \cite{dagan2}. Reinstating the superconductor
characteristic velocity in (\ref{zero-T-delta}) through the change
$e_{i}B\rightarrow(\hbar v^{2}/c)e_{i}B$, using $\hbar v \simeq
0.25 eV \AA$ for YBCO, and $e_{i} = e$ for one of the nodal
directions we find $\alpha\simeq 0.13$. It can be checked that
this result is not in contradiction with the dropping of the cubic
term in (\ref{gapeq2}).

We underline that the solution (\ref{zero-T-delta}) is the only
extremal at zero temperature. That is, $\Delta^{(0)}_{i}=0$ is not
even an extremal of the free-energy at $T=0$. Moreover, the
breaking of the symmetry by the magnetic field at $T=0$ occurs
independently of the strength of the Yukawa coupling $g$,
\textit{i.e.} there is no critical $g$ in this case.

A similar zero temperature behavior was previously found in our
studies of a (3+1)-dimensional HY model in the presence of a
magnetic field \cite{efavian1,efavian2}. However, there is a
fundamental difference between that case and the one we are
considering here. In the model of Refs. \cite{efavian1,efavian2},
the magnetic field breaks a discrete chiral symmetry producing a
chiral condensate $\langle\overline{\Psi}\Psi\rangle$. Such a
symmetry breaking is a manifestation of the phenomenon, introduced
by Gusynin, Miransky, and Shovkovy \cite{mira-gus-sho}, of
magnetic catalysis of chiral symmetry breaking (MCCSB)
\cite{klimenko}. The MCCSB mechanism has been suggested as a
possible explanation for several transport properties of strongly
correlated electronic systems \cite{transp-properties}.
Nevertheless, in the $id_{xy}$ theory considered in the present
paper, the condensation phenomenon has a different physical
character, since the Lagrangian (\ref{fouraction}) has explicitly
broken chiral symmetry due to the HY term. As it happens in MCCSB,
here too the fermion infrared dynamics in the presence of a
magnetic field induces the appearance of a gap and a
fermion-antifermion condensate, but in this case it is a
parity-breaking condensate $\langle
\overline{\Psi}\bigwedge\Psi\rangle$, instead of the chiral
condensate $\langle\overline{\Psi}\Psi\rangle$ associated to the
MCCSB. We call this new mechanism of magnetic-field-induced
dynamical symmetry breaking: magnetic catalysis of parity symmetry
breaking. Physically, the parity condensate corresponds to the
induction of a magnetic moment by the magnetic field. The
induction of a magnetic moment was proposed by Laughlin
\cite{laugh} as the key to explain the odd behavior of the thermal
conductivity at $B\neq 0$ in high-$T_{c}$ superconductors
\cite{krisha, ando1}.

As it is known, thermal effects tend to erase the generated
condensates. In the present case, this is reflected in the
tendency of the gap solution to vanishing
($\Delta_{i}(T)\rightarrow 0$) as $T$ is increased. That is, we
find that at some critical temperature $T_{ci}$, that depends on
the magnetic field, a second order phase transition takes place
from a gapped to a gapless state for the $i^{th}$ nodal direction,
independently of what occurs in the other nodal direction. Near
the transition temperature $T_{ci}$ the gap solution can be
written as
\begin{equation}\label{t-delta}
  \Delta_{i}(T)=\frac{4}{\alpha T}(T^{2}_{ci}-T^{2})
\end{equation}
with $T_{ci}=\alpha\sqrt{|e_{i}B|}=(\sqrt{\alpha/8})
\Delta^{(0)}_{i}$. Using the gap $\Delta^{(0)}_{i}=4 meV$, which
corresponds to a field of $15T$ in the $600 \AA$ film of
underdoped (1,1,0)-oriented YBCO, measured in decreasing field
\cite{dagan2}, we estimate a critical temperature $T_{c}\simeq
6K$.

Finally, we would like to comment on the anisotropic behavior of
the proposed model in the presence of a magnetic field. As seen
from Eq. (\ref{zero-T-delta}), the induced gap $\Delta_{i}(T)$ and
the critical temperature $T_{ci}$ of the $i^{th}$ nodal direction
depend on the strength of the effective charge $e_{i}$ of the
quasiparticles of that particular direction. The smaller the
effective charge $e_{i}$, the smaller the $\Delta_{i}(T)$ and the
$T_{ci}$. Then, given a magnetic field $B$, we have that for
$T_{c2}(B)\leq T \leq T_{c1}(B)$ no ZBCP splitting will be
observed if the film is oriented along direction $2$, but it will
be present for films oriented along $1$. In this sense, it would
be interesting to have available data of ZBCP measurements taken
along the nodal direction perpendicular to the one used in
\cite{dagan2}, as in that case one could use the relation
$e_{2}=(\frac{T_{c2}}{T_{c1}})^{2}e_{1}$ to estimate the ratio of
the two effective charges. We think that this anisotropic
condensation mechanism is also the basis to explain the
anisotropic transport properties in underdoped superconducting
samples \cite{krisha,ando1,ando2,ando3}.

This research was supported by the National Science Foundation
under Grant No. PHY-0070986.


\begin{thebibliography}{}
\bibitem{votja1}M. Votja, Y. Zhang, and S. Sachdev, Phys. Rev. B \textbf{62}
6721 (2000).
\bibitem{votja2}M. Votja, Y. Zhang, and S. Sachdev, Phys. Rev. Let. \textbf{85}
4940 (2000).
\bibitem{valla}T. Valla \textit{et al}., Science \textbf{285}, 2110 (1999).
\bibitem{corson}J. Corson \textit{et. al}, Phys. Rev. Lett. \textbf{85}, 2569 (2000).
\bibitem{quinlan} S. M. Quinlan, D. J. Scalapino, and N. Bulut, Phys. Rev. B
\textbf{49} 1470 (1994); M. L. Titov, A. G. Yashenkin, and D. N.
Aristov, \textit{ibid} \textbf{52},10626 (1995).
\bibitem{khvesh}D. V. Khveshchenko, and J. Paaske, Phys. Rev.
Lett. \textbf{86}, 4672 (2001).
\bibitem{b-tunneling}M. Covington \textit{et al}., Phys. Rev. Lett. \textbf{79}, 277 (1997);
M. Aprili, E. Badica, and L. H. Greene, Phys. Rev. Lett.
\textbf{83}, 4630 (1999); R. Krupke and G. Deutscher, Phys. Rev.
Lett. \textbf{83}, 4634 (1999).
\bibitem{hu}C. R. Hu, Phys. Rev. Lett. \textbf{72} 1526 (1994); S.
Kashiwaya \textit{et. al}, Phys. Rev. B \textbf{51} 1350 (1995).
\bibitem{dagan1} Y. Dagan and G. Deutscher, Phys. Rev. Lett. \textbf{87} 177004 (2001).
\bibitem{fogels} M. Fofgelstr\"{o}m, D. Rainer, and J. A. Sauls, Phys. Rev. Lett. \textbf{79}, 281 (1997)
\bibitem{dagan2} Y. Dagan and G. Deutscher, Phys. Rev. B \textbf{64} 092509 (2001);
R. Beck \textit{et. al}, cond-mat/0212447.
\bibitem{laugh} R.B. Laughlin, Phys. Rev. Lett. \textbf{80} 5188
(1998).
\bibitem{krisha} K. Krishana \textit{et al.}, Science
\textbf{277}, 83 (1997).
\bibitem{ando1} Y. Ando \textit{et. al}, Phys. Rev. Lett. \textbf{88} 147004 (2002).
\bibitem{ando2} Y. Ando \textit{et. al}, Phys. Rev. Lett. \textbf{88} 137005 (2002).
\bibitem{ando3} Y. Ando, A. N.
Lavrov, and S. Komiya, Phys. Rev. Lett. \textbf{90} 247003 (2003).
\bibitem{lee}P. A. Lee and X.-G. Wen, Phys. Rev. Lett. \textbf{78} 4111
(1997).
\bibitem{annett} J. Annett, N. Goldenfeld, and A. J. Leggett,
Physical Properties of High Tc Superconductors, edited by D.M.
Ginsberg (World Scientific, Singapore, 1996), Vol.5.
\bibitem{efavian1}E.J. Ferrer and V. de la Incera,
Phys.Lett.B\textbf{ 481},287 (2000); V. de la Incera, in Proc. of
the Intern. Conf. on Quantization, Gauge Theory, and Strings,
V.II, pag. 316, (ed. A. Semikhatov, M. Vasiliev, V. Zaikin)
Scientific World 2001.
\bibitem{efavian2} E. Elizalde, E.J. Ferrer and V. de la
Incera, hep-ph/0209324 (Phys. Rev. D in press).
\bibitem{mira-gus-sho} V. P. Gusynin, V. A. Miransky, and I. A. Shovkovy,
Phys. Rev. Lett. \textbf{73}, 3499 (1994); Phys. Lett. B {\bf
349}, 477 (1995); Phys. Rev. D \textbf{52}, 4747 (1995); Nucl.
Phys. B \textbf{462}, 249 (1996); \textit{ibid} \textbf{563}, 361
(1999).
\bibitem{klimenko} K.G. Klimenko, Z. Phys. C \textbf{54}, 323 (1992); D.-S Lee, C. N. Leung, and Y. J. Ng, Phys. Rev D
\textbf{55}, 6504 (1997).
\bibitem{transp-properties} G.W. Semenoff, I. A. Shovkovy, and
L.C.R. Wijewardhana, Mod. Phys. Lett. A \textbf{13}, 1143 (1998);
W.~V. Liu, Nucl. Phys. B {\bf 556}, 563 (1999); K. Farakos and
N.E. Mavromatos, Phys. Rev. B \textbf{57}, 3017 (1998);
E.J.Ferrer, V.P. Gusynin, and V. de la Incera, Mod. Phys. Lett.
B{\bf 16}, 107 (2002); Eur.Phys.J.B \textbf{33}, 397 (2003); V.P.
Gusynin and V. A. Miransky, cond-mat/0306143.


\end{thebibliography}
\end{document}